# Optimizing Federated Learning Configurations for MRI Prostate Segmentation and Cancer Detection: A Simulation Study

Ashkan Moradi[1], Fadila Zerka[1], Joeran Sander Bosma[2], Mohammed R. S. Sunoqrot[1,6], Bendik S. Abrahamsen[1], Derya Yakar[3,4], Jeroen Geerdink[5], Henkjan Huisman[1,2], Tone Frost Bathen[1,6], Mattijs Elschot[1,6]

[1]Department of Circulation and Medical Imaging, Norwegian University of Science and Technology, Trondheim, Norway

[2]Diagnostic Image Analysis Group, Department of Medical Imaging, Radboud University Medical Center, Nijmegen, The Netherlands

[3]Medical Imaging Center, Departments of Radiology, University of Groningen, University Medical Center Groningen, Groningen, The Netherlands

[4] Netherlands Cancer Institute, Department of Radiology, Amsterdam, The Netherlands

[5]Department of Information and Organization, Hospital Group Twente, Almelo, The Netherlands

[6]Department of Radiology and Nuclear Medicine, St. Olavs Hospital, Trondheim University Hospital, Norway

Correspondence: ashkan.moradi@ntnu.no

**Summary statement:** Federated learning enables collaborative model training, enhancing the performance and generalizability of trained models for MRI-based prostate gland segmentation and clinically significant prostate cancer detection compared with locally trained models.

**Key Points:**

- A retrospective study was conducted to evaluate federated learning (FL) for collaborative prostate segmentation (n = 1294 patients) and clinically significant prostate cancer (csPCa) detection (n = 1440 patients) using biparametric MRI.

- Compared with the average performance of the clients, the FL-baseline model demonstrated higher performance in prostate segmentation (Dice score increased from $0.73 \pm 0.06$ to $0.87 \pm 0.03$; $P \leq 0.01$) and lesion detection (detection score, defined as the average of the area under the receiver operating characteristic curve and average precision, increased from $0.63 \pm 0.07$ to $0.72 \pm 0.06$; $P \leq 0.01$).

- Optimizing the FL configuration improved the csPCa detection performance compared to the FL-baseline model (detection score increased from $0.72 \pm 0.06$ to $0.74 \pm 0.06$; $P \leq 0.01$), while no significant improvement was observed for prostate segmentation (Dice scores: $0.87 \pm 0.03$ vs. $0.88 \pm 0.03$; $P > 0.05$).

**Abbreviations:** csPCa = clinically significant prostate cancer, bpMRI = biparametric MRI, DL = deep learning, FL = federated learning, CL = centralized learning, PI-CAI = Prostate Imaging: Cancer Artificial Intelligence, AUC = area under the receiver operating characteristic curve, AP = average precision

**Keywords:** Federated learning, prostate cancer, magnetic resonance imaging, cancer detection, deep learning.

**ABSTRACT**

**Purpose:** To develop and optimize a federated learning (FL) framework across multiple clients for biparametric MRI prostate segmentation and clinically significant prostate cancer (csPCa) detection.

**Materials and Methods:** A retrospective study was conducted using Flower FL to train a nnU-Net-based architecture for MRI prostate segmentation and csPCa detection, using data collected from January 2010 to August 2021. Model development included training and optimizing local epochs, federated rounds, and aggregation strategies for FL-based prostate segmentation on T2-weighted MRIs (four clients, 1294 patients) and csPCa detection using biparametric MRIs (three clients, 1440 patients). Performance was evaluated on independent test sets using the Dice score for segmentation and the Prostate Imaging: Cancer Artificial Intelligence (PI-CAI) score, defined as the average of the area under the receiver operating characteristic curve and average precision, for csPCa detection. P-values for performance differences were calculated using permutation testing.

**Results:** The FL configurations were independently optimized for both tasks, showing improved performance at 1 epoch 300 rounds using FedMedian for prostate segmentation and 5 epochs 200 rounds using FedAdagrad, for csPCa detection. Compared with the average performance of the clients, the optimized FL model significantly improved performance in prostate segmentation (Dice score increase from 0.73±0.06 to 0.88±0.03; P≤0.01) and csPCa detection (PI-CAI score increase from 0.63±0.07 to 0.74±0.06; P≤0.01) on the independent test set. The optimized FL model showed higher lesion detection performance compared to the FL-baseline model (PI-CAI score increase from 0.72±0.06 to 0.74±0.06; P≤0.01), but no evidence of a difference was observed for prostate segmentation (Dice scores, 0.87 ± 0.03 vs 0.88 ± 03; P>0.05).

**Conclusion:** FL enhanced the performance and generalizability of MRI prostate segmentation and csPCa detection compared with local models, and optimizing its configuration further improved lesion detection performance.

## 1 INTRODUCTION

Owing to the high global mortality rate of prostate cancer (PCa), the early detection of clinically significant PCa (csPCa), i.e., with a Gleason grade 2 or higher, is crucial for improving patient survival rates. International guidelines recommend MRI before prostate biopsies [1]; along with evaluating the likelihood of csPCa based on the Prostate Imaging-Reporting and Data System, radiologists typically segment the prostate gland for the calculation of prostate-specific antigen (PSA) density and to guide ultrasound fusion during the targeted biopsy. The increasing number of MRI examinations [2] has led to the use of deep learning (DL) algorithms to reduce the workload of radiologists in both gland segmentation and cancer detection [3]; however, larger and more diverse datasets than those typically available within a single institution are needed to train robust and generalizable DL algorithms. Consequently, sharing of patient data among various institutions is crucial for collaborative learning, which exposes sensitive information to potential threats.

Federated learning (FL), which facilitates decentralized model training, has emerged as a promising approach to address these concerns. Using this method, clients train their models locally and share only the model parameters with a trusted server for aggregation with parameters from other clients. This framework ensures data privacy while enabling collaborative training and is particularly attractive for digital health applications because it does not necessitate data centralization from participating clients. Although challenges are inherent in FL, including data heterogeneity and privacy risks, this framework has been effectively utilized in healthcare for medical image segmentation and cancer detection [4-8]. Kades et al. [9] outlined a framework for medical data analysis in real-world clinical FL environments, which is implemented using Kaapana, an open-source toolkit for artificial intelligence-driven medical data analysis that enables

FL by adapting nnU-Net to an FL environment [10, 11]. In the field of prostate research, FL has been utilized for both MRI-based prostate gland segmentation [12, 13] and PCa detection [14, 15]. However, despite considerable progress in this field, limited studies on prostate imaging have comprehensively evaluated the impact of optimizing FL configurations on prostate gland segmentation and lesion detection based on biparametric MRI (bpMRI) data, comprising T2-weighted (T2W) and diffusion-weighted imaging.

The primary aim of this study was to investigate the impact of data federation through FL on the performance of bpMRI-based prostate segmentation and csPCa detection. Additionally, we explored the effects of various combinations of local epochs, federated rounds, and server-side aggregation strategies, on the performance of prostate segmentation and csPCa detection models.

## MATERIALS AND METHODS

### Patients and Data

This study involved a comprehensive investigation of the implementation and optimization of FL for biparametric MRI-based prostate segmentation and csPCa detection. For the prostate segmentation task, FL scenarios were explored across four clients, each using T2W MRIs with corresponding manually annotated prostate segmentations. Clients S1, S3, and S4 utilized the following publicly available datasets: Prostate158 [16], N = 138; Prostate-MRI-US-Biopsy [17-19], N = 763; and PROSTATEx [20], N = 282. Although these public datasets contain a higher number of patients than those listed above, for this study, only T2W sequences with available manual segmentations were included. The in-house multi-parametric MRI dataset utilized in this study was previously collected as part of a prospective study [21], from which client S2 utilized patients with available manual T2W MRI segmentations from 111 patients as part of a retrospective analysis. These cases were utilized in part by Patsanis et al. [22] to segment the 3D prostate volumes at a single center; however, in this study, the focus was on FL segmentation. These patients underwent examinations at St. Olavs hospital, Trondheim University Hospital, Trondheim, Norway between November 2014 and December 2017 for suspicion of PCa. The Regional Committee for Medical and Health Research Ethics approved the use of the dataset (identifier REK 2017/576) and all patients provided informed consent before the start of the prospective study [21]. Whole prostate volumes were manually delineated by a radiologist with more than a decade of experience, and subsets of 10 patients were randomly selected from each client's training set, serving as local test sets to assess the performance of the trained models. The remaining data were used as 80/20 training/validation sets for each client, while the PROMISE 12 training set [23] (N = 50 patients) was used as an independent test set to evaluate the trained models.

For the csPCa detection task, a scenario was investigated involving three clients with bpMRI data from the public training and development datasets of the Prostate Imaging: Cancer Artificial Intelligence (PI-CAI) Grand Challenge [24], compiled by three Dutch institutes, clients D1, D2, and D3 with 350, 800, and 350 patients, respectively. The performance of each client's trained models was evaluated using subsets of 20 randomly selected patients from the training sets, serving as local test sets. The remaining data were used as 80/20 training/validation sets for each client. Additionally, an independent internal test set, used as part of the hidden test set in the PI-CAI Grand Challenge and comprising 199 patients, was utilized to evaluate the trained models. The patient data in this test set were collected as part of the same study as the in-house prostate gland segmentation task dataset from client S2. In the csPCa test set, lesions were confirmed via histopathology reports, and the absence of csPCa in patients without a biopsy or with negative biopsy results was confirmed through a 3-year follow-up.

**Method and Model Training**

For each client, MRI-based prostate segmentation and csPCa detection tasks were performed using nnU-Net [10], which automatically configures a U-Net-based pipeline that is tailored to the specific geometry of the provided data and available computational resources. This process results in the implementation of local models, in which clients are trained separately using only their own data. A centralized learning (CL) scenario was also executed, in which data from all of the clients are collected in one location and training is performed using all available data. To ensure convergence of the models, the local and CL model training were each iterated for 300 epochs for prostate segmentation and 1000 epochs for csPCa detection. For FL implementation, each client machine trained a local model and transmitted the model parameters to a trusted server for aggregation. The Flower FL framework [25] was utilized to implement the FL models, the details of which are available in the supplemental material. The code is publicly available on GitHub.

To ensure a fair comparison between the CL and FL models, the training and validation sets used for the CL-baseline were composed of the combined training and validation sets from all of the contributing clients. This process ensured identical conditions for training both the CL and FL models, allowing for a direct comparison of their performance. An overview of the network topologies, illustrated in Figures 1 and 2, shows that FL training began with clients connecting to the network and independently initiating training on their local data. Each client machine was hosted in a secure local cloud environment with adequate resources for training a U-Net model. For prostate segmentation, each client utilized T2W images as the input data and trained its local model to produce a voxel-level binary map that determined whether each voxel was part of the prostate. For lesion detection, each client used bpMRI data, including T2W images, high b-value diffusion-weighted images, and apparent diffusion coefficient maps, as the input data and trained its local model to generate a voxel-level probability map that indicated the likelihood of each voxel being cancerous or non-cancerous. These detection maps were then post-processed to extract lesions and predict the patient-level likelihood of harboring csPCa [26, 27]. After (E) epochs of local training, the model parameters were shared with a trusted server that utilized the FedAvg algorithm [28] to aggregate the model parameters into a global model. The global model was then returned to the clients to initiate another round of local training, and this iterative process continued for a total of (R) rounds to refine the global model. The local epoch (E) was defined as the number of client iterations on mini-batches and the federated round (R) as the number of model aggregations at the server.

The structural hyperparameters of FL were then optimized, focusing on inter- and intra-client iterations and aggregation strategies rather than local training algorithms. Specifically, the epoch-round combinations were optimized by evaluating various epoch and round combinations in FedAvg FL training, while keeping E × R constant to ensure model convergence. The combination that exhibited the highest performance on the combined validation sets was identified as the optimal model. Similarly, E-R was fixed and various aggregation strategies, including FedAvg, FedAdagrad [29], FedAdam [29], FedYogi [29], and FedMedian [30], were evaluated to identify the server-side strategy that delivered the best performance on the combined validation set. The performances of the local models, CL- and FL-baseline models, and optimized FL scenarios were then compared. In the FL implementations, limitations in time and computational resources restricted the selection of the optimal parameters to those based on training a single model (one-fold cross-validation) and their evaluation using a combined validation set of aggregated data from all clients. The final test performance of the selected FL model, alongside other baseline models, is presented as an ensemble of the five models resulting from five-fold cross-validation. Additional details on the selection of the optimization parameters are provided in the supplemental material.

**Statistical Analysis**

Prostate gland segmentation models were evaluated by computing the Dice score, average 95% Hausdorff distance (HD95), and average relative volume difference across various test sets. For the csPCa detection task, the PI-CAI Grand Challenge [24] guidelines were followed, the area under the receiver operating characteristic curve (AUC) was utilized to assess the patient-level diagnostic performance, and the average precision (AP) was used to evaluate lesion-level detection performance. The AP summarizes the precision-recall curve as the weighted mean of precisions achieved at each decision threshold. The overall performance of each model for csPCa detection was then evaluated using the PI-CAI score, which is calculated as follows: (AUC + AP) / 2. This score combines both patient- and lesion-level performances. Bootstrapping with a sample size of 1000000 was used to calculate the 95% confidence intervals (CIs) of the performance metrics. To determine the probability of one model outperforming another, we performed a permutation test on the performance metric over five differently trained models using 1000000 iterations with a statistical significance threshold of 0.05. Statistical analyses were implemented in Python 3.9.

**RESULTS**

**Prostate Gland Segmentation**

The prostate gland segmentation task utilized data from 1294 male patients across four clients, with both local and CL-baseline models trained for E = 300 epochs. The FL-baseline model used E = 1 local epoch with R = 300 federated rounds and FedAvg as the server-side aggregation strategy. The models with fewer local epochs and a larger number of federated rounds demonstrated superior Dice scores. The optimal epoch-round combination occurred during the FL-baseline experiment at E = 1 and R = 300. For the server-side aggregation strategy, the FL-baseline scenario was repeated with modifications made solely to the aggregation strategy to train various models. The model that utilized the FedMedian aggregation strategy demonstrated the highest Dice score and was considered the optimal model. Further details on the data distribution and optimization process are provided in the supplemental material.

The five-fold cross-validated Dice scores are provided in Table 1 for the local, CL-baseline, FL-baseline, and optimized FL models evaluated using various test sets. The FL-tuned refers to the performance of the model with an optimized E-R combination (E = 1, R = 300) and FedMedian as the server-side aggregation strategy. FL consistently outperformed local models and matched the performance of the CL-baseline model (Tabel 1). Table 2 presents the Dice score, 95% Hausdorff distance, and relative volume distance metrics for the different models tested using the PROMISE12 test set. The FL-baseline model showed an enhanced performance compared with local models, improved the Dice score from $0.73 \pm 0.06$ (average of local models) to $0.87 \pm 0.03$ ($P \leq 0.01$), and closely matched the CL-baseline model for all metrics. Additional tuning of the FL model showed no evidence of improvement in the segmentation task (Dice scores, $0.87 \pm 0.03$ vs $0.88 \pm 03$; $P > 0.05$). The supplemental material shows an analysis of model performance at specific Dice score thresholds, offering further insights into potential clinical applications. Examples of MRI-based prostate gland segmentation results for various models in conjunction with the ground truth are shown in Figure 3; the FL model achieved highly accurate segmentation in cases where local models failed to accurately segment the prostate gland.

**Prostate Lesion Detection**

The csPCa detection task utilized 1500 bpMRI scans for male patients (mean age: 65.6±7.2 years range 35-92) from three clients. Both local and CL-baseline models were trained for E = 1000 epochs, while the FL-baseline model was trained for E = 1 local epoch across R = 1000 federated rounds using FedAvg aggregation. During optimization, the models were trained using various combinations of epochs and rounds, with each combination evaluated using the combined validation set. The extraction of patient- and lesion-level metrics from the voxel-level predictions of the trained model followed the PI-CAI guidelines, a summary of which is provided in the supplemental material. After post-processing, the optimal combination that achieved the highest PI-CAI score consisted of a small number of local epochs (E = 5) and a relatively high number of federated rounds (R = 200). To optimize the aggregation strategy, we repeated the FL-baseline scenario using different server-side aggregation strategies for training various models, with the optimal strategy identified as FedAdagrad aggregation. Further details on the data distribution and the optimization process are available in the supplemental material and in the article by Moradi et al. [31].

Table 3 presents the five-fold cross-validated PI-CAI scores of the csPCa detection models evaluated on various test sets. We examined the impact of independently optimizing the E-R combination and the aggregation strategy in FL-E5R200 and FL-FedAdagrad, respectively, as well as that of using the optimized values for both in the FL-tuned model. Similar to the segmentation task, the FL-baseline consistently improved the PI-CAI score compared to local models, from $0.63 \pm 0.07$ (average of local models) to $0.72 \pm 0.06$ (P≤0.01), and closely matched the CL-baseline performances. Additional evaluation metrics for the internal test set are provided in Table 4, which confirm the findings presented in Table 3 and show that performance can be further refined by fine-tuning the FL parameters (PI-CAI score increased from $0.72 \pm 0.06$ to $0.74 \pm 0.06$; $P \leq 0.01$). For additional clinical insights, details on the average number of false-positive lesions per patient at specific sensitivity levels are provided in the supplemental material.

The receiver operating characteristic and precision-recall curves presented in Figure 4 confirm that the AUC and AP of the FL-baseline model were significantly higher than those for the local models, AUC from $0.86 \pm 0.06$ (average of local models) to $0.89 \pm 0.06$ ($P \leq 0.01$) and AP from $0.41 \pm 0.1$ (average of local models) to $0.56 \pm 0.1$ ($P \leq 0.01$), and were comparable to those of the CL-baseline model. Figure 4 also shows that the AUC and AP of the optimized FL models were higher than those of the FL-baseline, AUC from $0.89 \pm 0.06$ to $0.91 \pm 0.04$ ($P \leq 0.01$) and AP from $0.56 \pm 0.1$ to $0.58 \pm 0.1$ ($P \leq 0.01$). Figure 5 shows ground truth lesions and model predictions as a qualitative performance evaluation of three patients from the in-house test set. In the first case, all models accurately identified the csPCa lesion, whereas in the second case, only the centralized and fine-tuned FL models succeeded in the task. In the third and more challenging case, the tumor extended beyond the prostate capsule, and only the fine-tuned FL model identified the csPCa lesion. These findings suggest that the performance of the fine-tuned FL model was comparable to that of the CL-baseline and surpassed that of locally trained models; additionally, the generalizability of the model was enhanced when evaluated using an independent test set. Figure 6, which illustrates the qualitative performance of MRI-based prostate segmentation and csPCa detection, shows that the performance of the FL-baseline model was comparable to that of the CL-baseline model, and further tuning of its configuration improved the detection rate of csPCa.

## DISCUSSION

This study investigated the effects of various combinations of epoch-rounds and server-side aggregation strategies on the performance of a federated learning (FL) framework for biparametric MRI (bpMRI)-based prostate segmentation and clinically significant prostate cancer (csPCa) detection. The baseline experiments showed that the performance of FL was comparable to that of centralized learning (CL) and better than that of local models on their respective test sets, suggesting that institutions can improve their model training for MRI-based prostate segmentation and csPCa detection through collaboration and without the need for data sharing. Furthermore, the performance improvement of the FL-baseline and optimized FL model evaluated on the independent test set demonstrated the generalizability of FL implementation compared with local models for both tasks. FL-baseline significantly improved ($P \leq 0.01$) prostate segmentation Dice ($0.73 \pm 0.06$ to $0.87 \pm 0.03$) and lesion detection score ($0.63 \pm 0.07$ to $0.72 \pm 0.06$) compared to the average of the local models.

Optimizing the combination of epoch-rounds and the aggregation strategy resulted in modest but consistent improvements in MRI-based csPCa detection across both the internal and external test sets compared with the FL-baseline model. Specifically, the csPCa detection model achieved superior results using E = 5, R = 200, and the FedAdagrad aggregation strategy. This performance was slightly better than that of the CL-baseline model, consistent with the findings on FL-based trained models reported by Kades et al. [9]. This difference may be attributed to the fact that the CL method does not consider case distribution across the contributing clients.

In evaluating the aggregation strategies, optimization-based approaches such as FedAdam, FedYogi, and FedAdagrad yielded better performance compared with FedAvg owing to their use of adaptive learning rates that can be adjusted based on historical gradient information. This allows for the improved management of client variability and faster convergence than that in the FedAvg [29]. These improvements, however, are dependent on additional fine-tuned hyperparameters. Although this study utilized only the values suggested in the literature [29], fine-tuning these parameters could improve the performance and convergence speed of these algorithms. Moreover, we speculate that combining a small number of local epochs with a relatively high number of federated rounds would result in improved performance, as this arrangement would allow clients to have sufficient local epochs for extracting information from their data while frequently aggregating to benefit from the information gained from other clients. When assessing the clinical impact of FL optimization, we found that tuning the FL configuration enhanced the performance of the model by increasing the number of test cases that achieved a specified Dice score threshold for MRI-based prostate segmentation and by reducing the average number of false-positive lesions per patient for csPCa detection.

Regarding the study limitations, although the FL infrastructure limits the exposure of patient-specific data to potential threats, it does not entirely address the privacy concerns associated with data sharing [32-34], as various attack strategies can still infer sensitive patient information through shared model parameters. Therefore, to enhance the protection of patient data, more sophisticated, privacy-preserving methods such as secure multi-party computation, differential privacy, and homomorphic encryption can be integrated with FL [33]. However, while these methods enhance privacy, they adversely affect performance, necessitating a consideration of the trade-off between privacy and performance [34]. Several studies on FL have suggested developing local algorithms, server-side aggregations, and client-server communications to enhance performance beyond the baseline scenario [35]. This study, however, focused on examining the impacts of fine-tuning the server-side aggregation strategies and epoch-round combinations on the performance of FL models. We utilized uniform preprocessing steps for all clients, which resulted in nnU-Net selecting the same model architecture, eliminating any compatibility challenges during aggregation. However, in real-world scenarios, variations in the preprocessing and computational resources utilized by each client can result in incompatibility during model aggregation. This issue can be addressed by having the server initiate the training process, imposing common model and preprocessing parameters on the clients. In real-world clinical environments, variations in data and available resources across institutions can result in differences in model performance and hinder the effective generalizability of these models. To account for this, we preserved the real-world data distribution in our simulation experiments; however, further challenges related to inter-institutional FL implementations might arise in real-world scenarios. Given the substantial time and computational resources required to evaluate five-fold cross-validated models for each parameter combination during the FL optimization, we selected optimal parameters based on the performance of a single trained model, as detailed in the supplemental material. The selected configurations were subsequently validated through the evaluation of five-fold cross-validated models using external test sets. In this study, some clients used a relatively small number of cases, which may have influenced optimization decisions due to the limited size of the validation set. Last but not least, although multiple open-source FL frameworks exist, all experiments in this study were performed using Flower FL, which may affect the generalizability of the findings.

In conclusion, our study showed that clients can improve model performance for MRI prostate segmentation and csPCa detection tasks using FL, and that fine-tuning FL configurations further improved performance for csPCa detection. This study is paving the way for the real-world implementation of FL and its practical application in detecting csPCa.

**Tables:**

*Table 1: The average Dice score of the 5-fold cross-validated prostate gland segmentation models assessed on various test sets.*

| Model/Test set | S1 | S2 | S3 | S4 | All10s | PROMISE12 |
|---|---|---|---|---|---|---|
| Local S1 | 0.35 | 0.87 | 0.64 | 0.86 | 0.68 | 0.47 |
| Local S2 | 0.91 | 0.93 | 0.89 | 0.91 | 0.91 | 0.82 |
| Local S3 | 0.88 | 0.89 | 0.91 | 0.89 | 0.89 | 0.84 |
| Local S4 | 0.90 | 0.91 | 0.89 | 0.92 | 0.91 | 0.79 |
| CL-baseline | 0.90 | 0.92 | 0.92 | 0.92 | 0.91 | 0.88 |
| FL-baseline | 0.90 | 0.92 | 0.92 | 0.91 | 0.91 | 0.87 |
| FL*-tuned | 0.90 | 0.93 | 0.91 | 0.92 | 0.91 | 0.88 |

Note.— Each local test set (S1, S2, S3, and S4) includes 10 randomly selected data samples, while All10s is the combined version of these sets with 40 patients and PROMISE12 is an independent test set with 50 patients. The FL*-tuned model represents the FL model with E=1 local epochs and R=300 federated rounds, employing FedMedian as the server-side aggregation strategy. The FL-baseline model utilizes E=1 local epochs and R=300 federated rounds with FedAvg as the server-side aggregation strategy, and local and CL-baseline models are iterated to E=300 epochs.

*Table 2: The average Dice score, Hausdorff 95% distance (HD95), and relative volume difference (RVD) scores of the 5-fold cross-validated prostate gland segmentation models assessed on the PROMISE12 test set with 50 patients.*

| Model | Dice | HD95 (mm) | RVD (%) |
|---|---|---|---|
| Local S1 | 0.47 (0.37,0.57) | 55.46 (47.83,62.82) | 49.77 (38.27,61.95) |
| Local S2 | 0.82 (0.77,0.86) | 15.28 (11.29,20.18) | 22.51 (16.09,31.65) |
| Local S3 | 0.84 (0.80,0.87) | 13.87 (10.79,17.46) | 19.47 (12.26,30.37) |
| Local S4 | 0.79 (0.73,0.84) | 12.47 (9.97,15.07) | 23.38 (17.30,30.31) |
| CL-baseline | 0.88 (0.86,0.90) | 9.14 (7.20,11.75) | 12.50 (8.46,18.32) |
| FL-baseline | 0.87 (0.85,0.89) | 11.20 (8.18,15.24) | 14.95 (9.68,23.39) |
| FL*-tuned | 0.88 (0.85,0.90) | 11.03 (7.96,15.07) | 14.20 (8.90,22.79) |

Note.— The FL*-tuned represents the FL model with E=1 local epoch and R=300 federated rounds, employing FedMedian as the server-side aggregation strategy. The FL-baseline model utilizes E=1 local epochs and R=300 federated rounds with FedAvg as the server-side aggregation strategy, and local and CL-baseline models are iterated to E=300 epochs. 95% confidence intervals of the metrics are included in parentheses, with a bootstrap sample size of b=1000000.

*Table 3: The PI-CAI scores, average of the area under the ROC curve (AUC) and average precision (AP), of the 5-fold cross-validated csPCa detection models assessed on various test sets.*

| Model/Test set | D1 | D2 | D3 | All20s | In-house |
|---|---|---|---|---|---|
| Local D1 | 0.91 | 0.74 | 0.52 | 0.73 | 0.66 |
| Local D2 | 0.98 | 0.76 | 0.60 | 0.76 | 0.60 |
| Local D3 | 0.68 | 0.71 | 0.65 | 0.65 | 0.63 |
| CL-baseline | 0.94 | 0.80 | 0.70 | 0.82 | 0.73 |
| FL-baseline | 0.94 | 0.82 | 0.73 | 0.82 | 0.72 |
| FL-E5R200 | 0.94 | 0.81 | 0.75 | 0.84 | 0.73 |
| FL-FedAdagrad | 0.93 | 0.85 | 0.71 | 0.84 | 0.74 |
| FL*-tuned | 0.94 | 0.87 | 0.67 | 0.84 | 0.74 |

Note.— Each local test set (D1, D2, and D3) includes 20 randomly selected data samples, while All20s is the combined version of these sets with 60 patients and the in-house test set is an independent set with 199 patients. The local epochs (E), federated rounds (R), and server-side aggregation strategies for the FL-based models are used as follows: FL*-tuned: E=5, R=200, and FedAdagrad; FL-baseline: E=1, R=1000, and FedAvg; FL-E5R200: E=5, R=200, and FedAvg; and FL-FedAdagrad: E=1, R=1000, and FedAdagrad. The local and CL-baseline models are iterated for E=1000 epochs.

*Table 4: Area under the ROC curve (AUC), average precision (AP), and PI-CAI scores, (AUC+AP)/2, of the 5-fold cross-validated csPCa detection models assessed on the in-house test set with 199 patients.*

| Model | AUC | AP | PI-CAI score |
|-------|-----|----|--------------| 
| Local D1 | 0.84 (0.78,0.89) | 0.48 (0.38,0.57) | 0.66 (0.59,0.72) |
| Local D2 | 0.88 (0.82,0.93) | 0.33 (0.23,0.44) | 0.60 (0.54,0.67) |
| Local D3 | 0.85 (0.80,0.90) | 0.41 (0.31,0.51) | 0.63 (0.57,0.70) |
| CL-baseline | 0.89 (0.85,0.94) | 0.57 (0.46,0.67) | 0.73 (0.66,0.80) |
| FL-baseline | 0.89 (0.83,0.94) | 0.56 (0.47,0.66) | 0.72 (0.66,0.79) |
| FL-E5R200 | 0.87 (0.82,0.93) | 0.60 (0.50,0.70) | 0.73 (0.67,0.81) |
| FL-FedAdagrad | 0.90 (0.86,0.95) | 0.59 (0.50,0.69) | 0.74 (0.69,0.81) |
| FL*-tuned | 0.91 (0.87,0.95) | 0.58 (0.48,0.67) | 0.74 (0.69,0.80) |

Note.— The local epochs (E), federated rounds (R), and server-side aggregation strategies for the FL-based models are used as follows: FL*-tuned: E=5, R=200, and FedAdagrad; FL-baseline: E=1, R=1000, and FedAvg; FL-E5R200: E=5, R=200, and FedAvg; and FL-FedAdagrad: E=1, R=1000, and FedAdagrad. The local and CL-baseline models are iterated for E=1000 epochs. 95% confidence intervals of the metrics are included in parentheses, with a bootstrap sample size of b=1000000.

**Figures:**

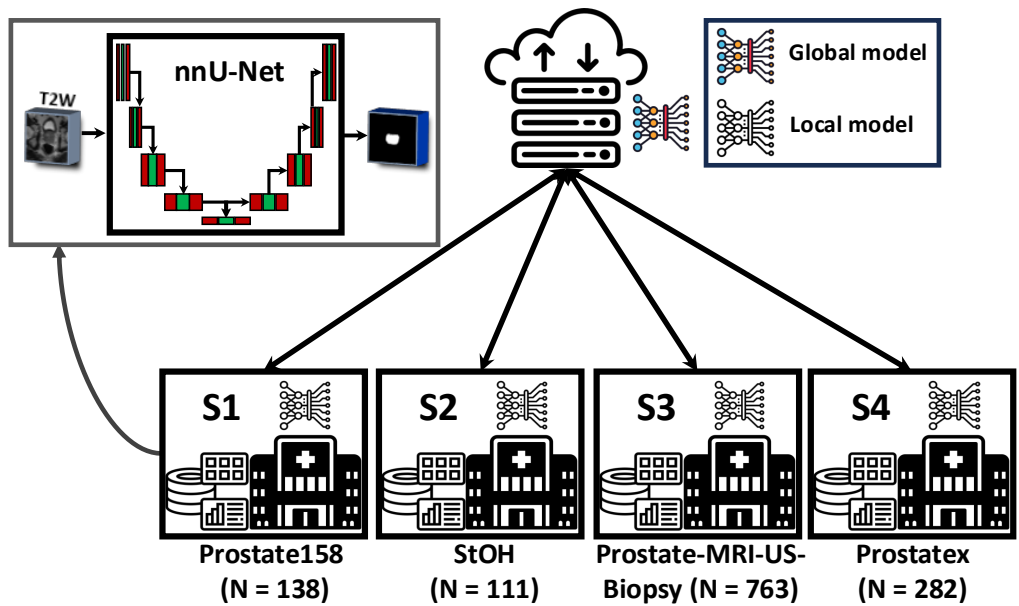

*Figure1 The federated learning (FL) network topology with 4 clients utilized for the prostate gland segmentation task. The FL-baseline model trains local models for E=1 epoch and subsequently shares the model parameters with the server for aggregation over R=300 rounds. Each client holds a dataset of T2-weighted (T2W) prostate MRIs: Prostate158 data with 138 male patient (mean age: 66.6±9 years range 35-84), St. Olavs Hospital (StOH) data with 111 male patient (median age = 64; range: 45–75 years) , Prostate-MRI-US-Biopsy data with 763 male patient, and PROSTATEx data with 282 male patient (mean age: 66 years range 48-83).*

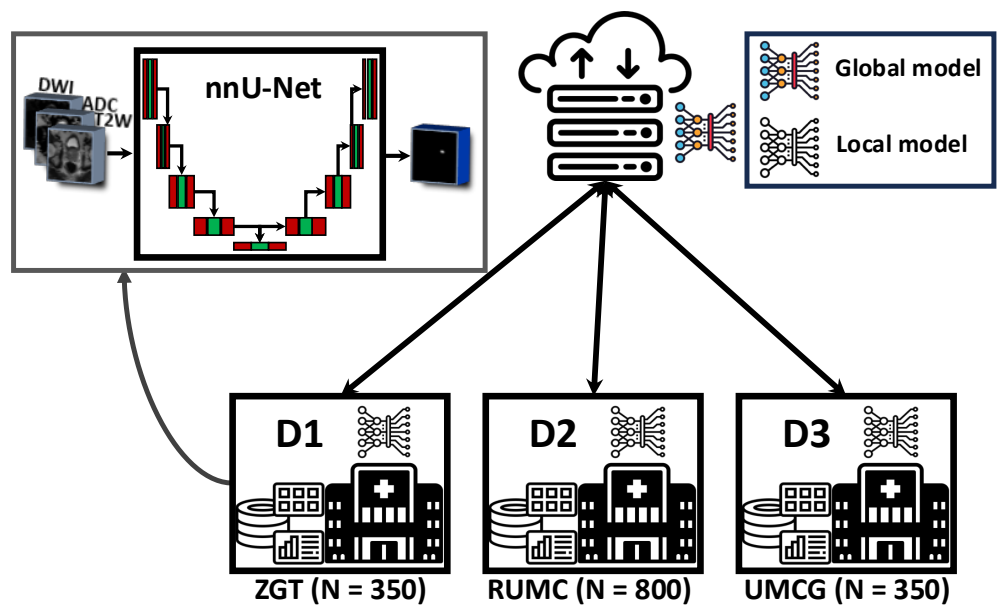

*Figure2 The federated learning (FL) network topology with three clients, D1, D2, and D3, utilized for the task of detecting clinically significant prostate cancer. The FL-baseline model trains local models for E=1 epoch and subsequently shares the model parameters with the server for aggregation over R=1000 rounds. Each client holds a dataset of prostate bpMRIs including T2-weighted (T2W) images, diffusion-weighted images (DWI), and apparent diffusion coefficient (ADC) maps: Ziekenhuis Groep Twente (ZGT) with 350 male patients (mean age: 66.6±7.4 years range 43–89), Radboud University Medical Center (RUMC) with 800 male patients (mean age: 64.6±7.1 years range 35–92), and University Medical Center Groningen (UMCG) with 350 male patients (mean age: 66.9±6.8 years range 45-83).*

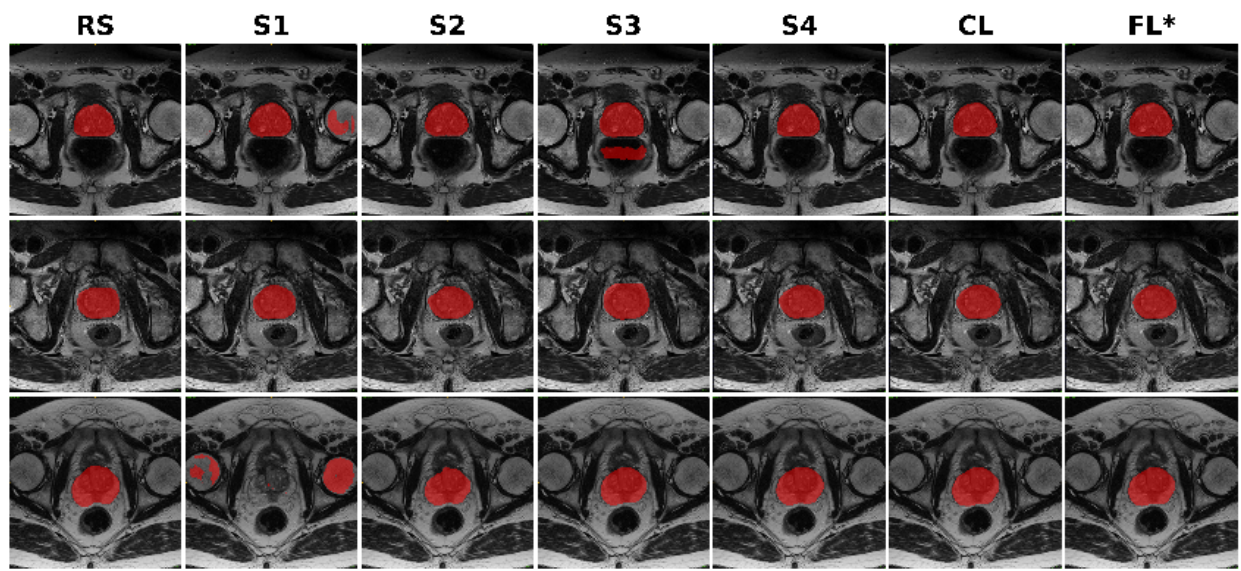

*Figure3 The qualitative performance representation of the prostate gland segmentation strategy is shown, with each row representing one test case, for fine-tuned FL (FL\*), CL-baseline (CL), and local models on a single test sample compared to the reference standard (RS). The test samples are selected to highlight the potential failures of the locally trained models, while demonstrating the consistent success of the CL and FL models in segmenting the prostate gland. The inability of the local model in S1 to accurately segment the prostate gland is noticeable in the second column of the figure. This level of failure for client S1 is also evident in **Table 1** and **Table2**, where its local performance is markedly inferior to that of other clients. This shows that the performance gain for S1 to join the federation is the highest, as its training dataset deviates the most from the other datasets in terms of image quality and input size.*

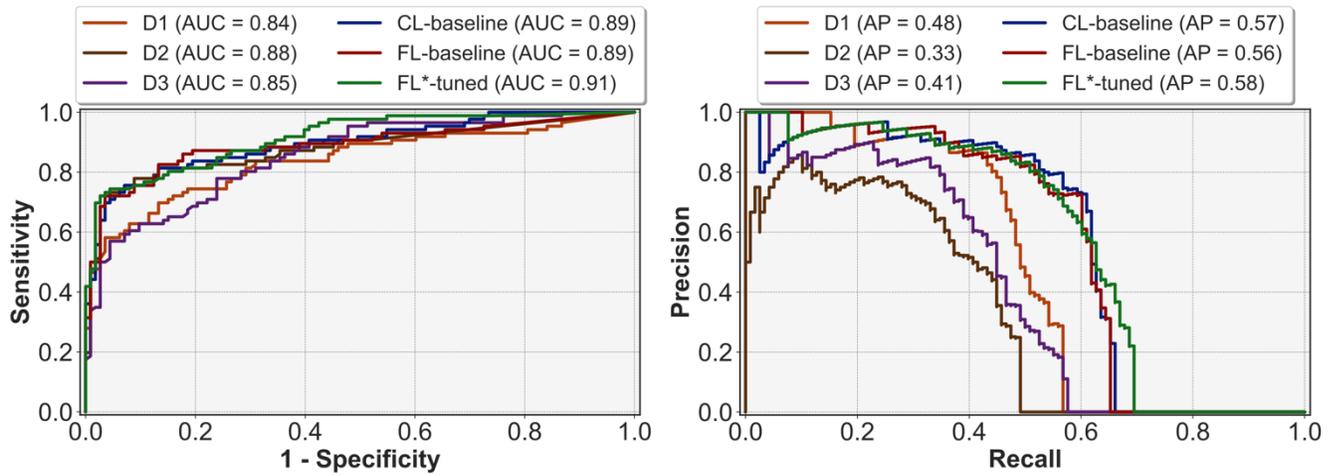

*Figure4 Receiver operating characteristic (ROC) and precision-recall (PR) curves for the 5-fold cross-validated models evaluated on the in-house test set with N=199 male patient data (mean age: 64±6.76 years range 44-76) for clinically significant prostate cancer detection. AUC = area under the ROC curve, AP = average precision*

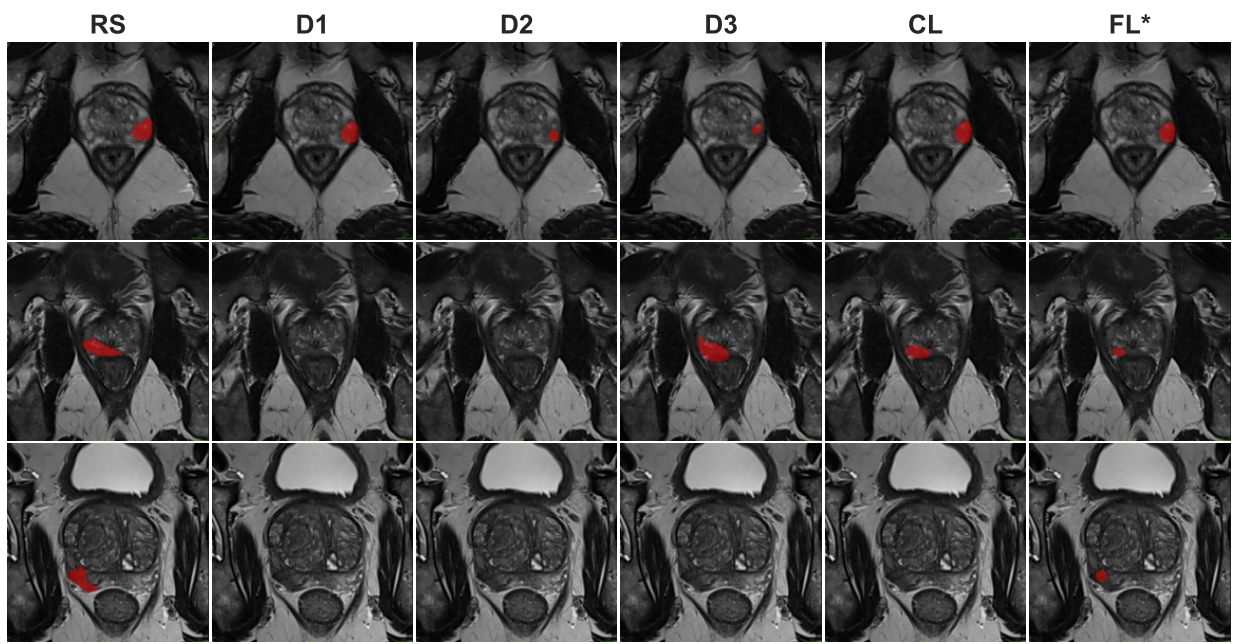

*Figure5 The qualitative performance representation of the clinically significant prostate cancer detection strategy is shown, with each row representing one test case. This includes fine-tuned FL (FL*), CL-baseline (CL), and local models, compared to the reference standard (RS). The test samples are selected to highlight the potential failures of the locally trained and centralized models, while demonstrating the consistent success of the fine-tuned FL model in detecting clinically significant prostate cancer.*

| RS | CL | FL | FL* |
|----|----|----|-----|

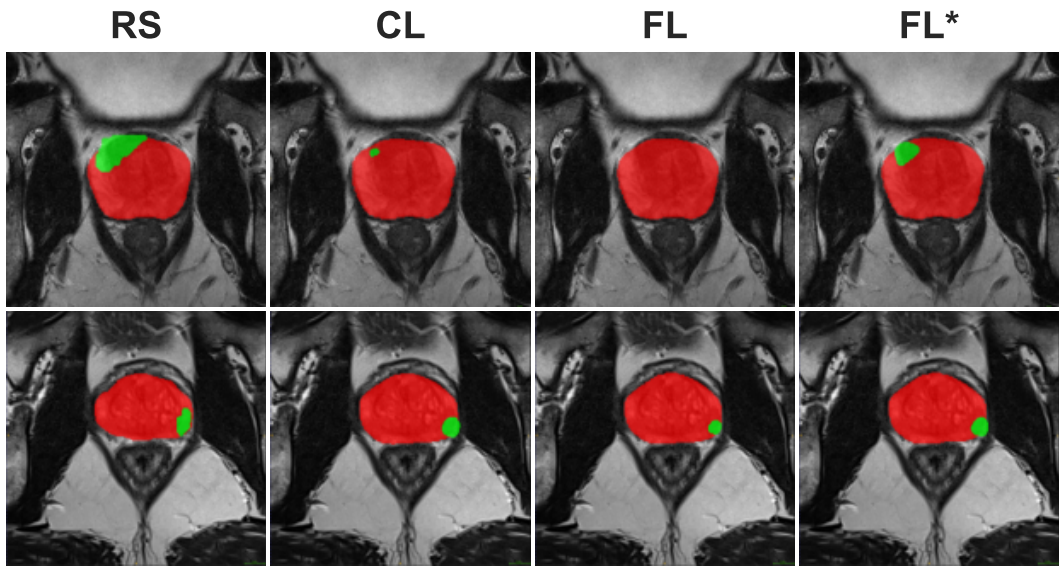

*Figure6 The qualitative performance representation of prostate gland segmentation and clinically significant prostate cancer detection strategies is presented for fine-tuned FL (FL*), FL-baseline (FL), and CL-baseline (CL) models, compared with the reference standard (RS). Each row represents a single test case, selected to highlight the high performance of both CL- and FL-based models in the whole prostate segmentation task. The first row demonstrates a patient where the FL baseline fails to detect clinically significant prostate cancer and fine-tuning the FL configuration improved performance.*